\documentclass[12pt]{article}
\usepackage{graphicx}
\usepackage[top=2cm, bottom=3cm, left=3cm, right=3cm]{geometry}

\title{The radial gradient of cosmic ray intensity in the Galaxy}
\author{$^{*}$A.D. Erlykin $^{1,2}$, A.W. Wolfendale $^{2}$ and V.A.Dogiel $^{1,3}$\\
$(1)$ P N Lebedev Physical Institute, Moscow 119991, Russia.\\
$(2)$ Physics Department, Durham University, Durham, DH1 3LE, UK\\
$(3)$ Moscow Institute of Physics and Technology,\\
 141700 Moscow Region, Dolgoprudnii, Russia}

\begin{document}
\maketitle

\footnote{$^{*}$Corresponding author: tel +74991358737 \\ 
 E-mail address: erlykin@sci.lebedev.ru}

\begin{abstract}
 The dependence of the cosmic ray intensity on Galactocentric distance is known to be 
much less rapid than that of the thought-to-be sources: supernova remnants. This is an 
old problem ('the radial gradient problem') which has led to a number of possible 
'scenarios'. Here, we use recent data on the supernova remnant's radial distribution 
and correlate it with the measured HII electron temperature ($T$). We examined two
models of cosmic ray injection and acceleration and in both of them the injection 
efficiency increases with increasing ambient temperature $T$. The 
increase is expected to vary as a high power of $T$ in view of the strong 
temperature-dependence of the tail of the Maxwell-Boltzmann distribution of particle 
energies. Writing the efficiency as proportional to $T^n$ we find $n\approx 8.4$. There
 is thus, yet another possible explanation of the radial gradient problem. 
\end{abstract}

\vspace{1mm}


\section{Introduction}
The manner in which cosmic ray (CR) particles are accelerated in the Galaxy and how 
they propagate is still the subject of argument although it is generally agreed that up
 to a few PeV or so supernova remnants (SNR) play an important role, as does the 
diffusion in the interstellar medium (ISM) \cite{Ginz}.  However, there are a few 
problems of detail among which one is the following: \\
why, when the surface density of SNR in the Galaxy falls so rapidly with 
Galactocentric radius, $R$, for $R > 3$kpc \cite{Case, Green}, does the CR intensity 
fall so slowly \cite{EW1,Stro1} ?

In the recent past we have examined this question in some detail \cite{EW4}, 
but here we study the problem again in the light of superior data on the radial 
gradient of SNR and the appreciation that in some models of particle injection into SN 
shocks the temperature of the ambient ISM has relevance \cite{Berez,Kang}.

\section{The input data for the analysis}
\subsection{The radial gradient of SNR}
The determination of the radial gradient is a matter of some difficulty on account of 
obscuration by dust and attenuation of the radio signals, by which distances are 
measured. Two, rather extreme, estimates have been made, by \cite{Case} and 
\cite{Green}. The first \cite{Case} peaks at $R\sim 4$kpc and drops to 40\% of its peak
 value by $\sim 10$kpc whereas the second \cite{Green} peaks at $R\sim 3$kpc and falls 
to 15\% by 10kpc. Insofar as the analysis in \cite{Green} is later and, \cite{EW4} 
points to shortcomings in the analysis of \cite{Case} we adopt the results from 
\cite{Green}. A representative surface density of SNR from \cite{Green} is given in 
Figure 1, denoted SNR.
\begin{figure}[hptb]
\begin{center}
\includegraphics[width=8cm,height=15cm,angle=-90]{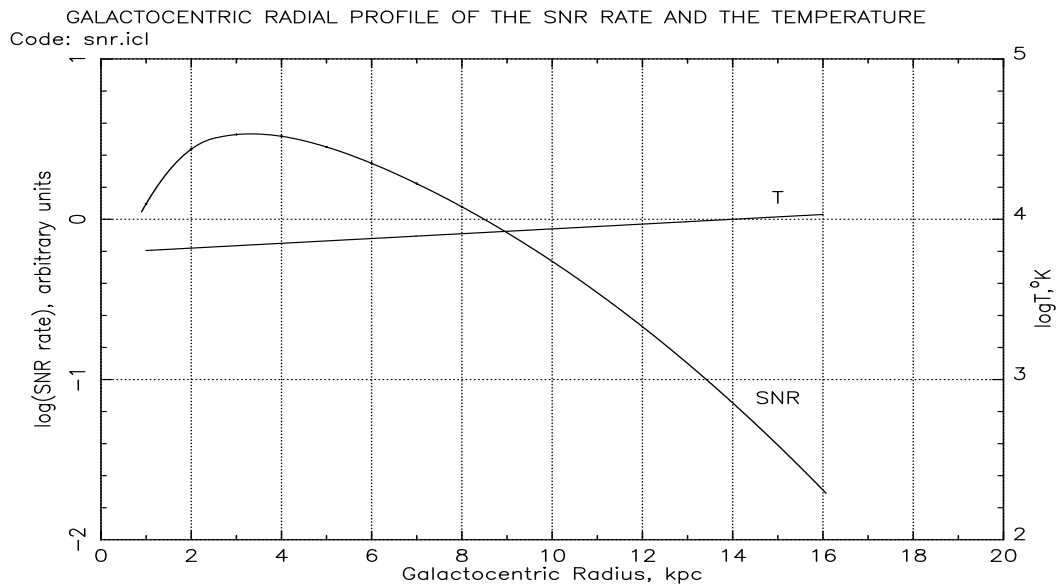}
\end{center}
\caption{\footnotesize Surface density of SNR vs. Galactocentric radius, $R$, [from 
\cite{Green}, Figure 4] (arbitrary scale of ordinate) and electron temperature, $T$ vs 
$R$ [from \cite{Quire}, Figure 5]}
\label{fig:fig1}
\end{figure}
\subsection{The radial distribution of CR}
  The analysis of the gamma-ray emissivity in the Galactic Disk derived from the COS-B 
data by \cite{Bhat} and \cite{Bloe1}  showed a flatter distribution of 
CRs than that of their presumed sources (SNR) as derived even from \cite{Case}. Later 
work by \cite{Bloe2,Digel,Stro2,Stro3}  as well as the emissivity from the Fermi-LAT 
\cite{Acke2,Tibal} came to the same conclusion (see also for a review of the history of
 the 'radial gradient' of the CR intensity \cite{Dogi1}). Here we use an amalgam and 
present the result in Figure 2.
\begin{figure}[hptb]
\begin{center}
\includegraphics[width=8cm,height=15cm,angle=-90]{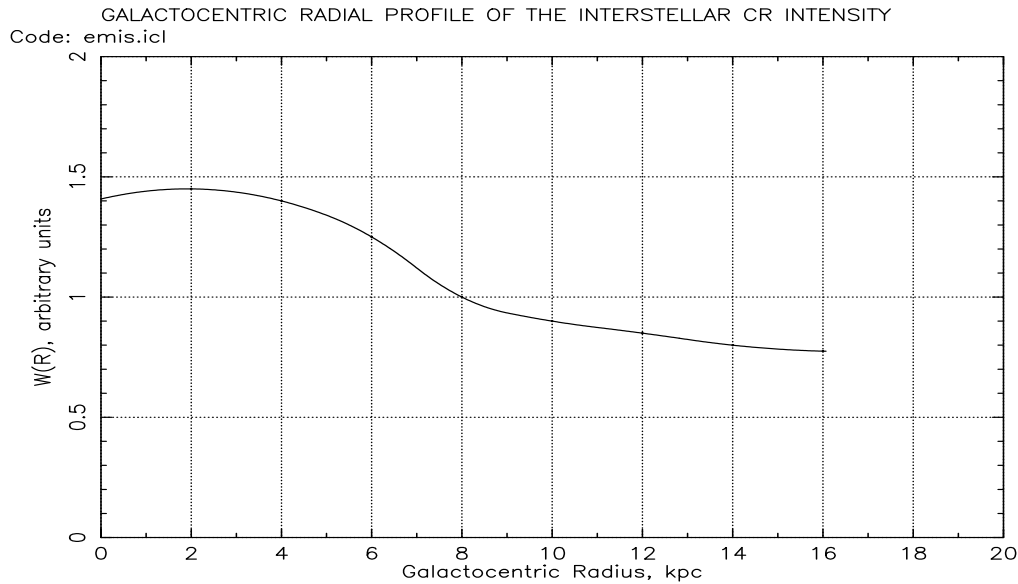}
\end{center}
\caption{\footnotesize Radial distribution of the CR intensity derived from gamma ray 
astronomy. See text for details.}
\label{fig:fig2}
\end{figure}
It can be remarked that the problem of the difference between the radial gradient (i.e.
 the dependence of CR density  on the Galactocentric radius $R$) of CR and sources is a
 live one. It seemed that a natural explanation  would be an effective spatial mixing 
of CRs due to the diffusion which produces a more or less uniform distribution of CRs 
in different parts of the Galactic Disk if it is surrounded by a giant halo. However, 
numerical calculations of \cite{Dogi1,Bloe2} for the free escape boundaries have shown,
 that even in the most favorable case of an extended halo, the diffusion model is 
unable to remove the signature of the source distribution even for the relatively 
smooth SNR distribution of \cite{Case}. The problem is even more aggravated for the 
sharper SNR distribution of \cite{Green}. One of the explanations was suggested by 
\cite{Breit} who assumed that CRs leave the disk faster from the regions of a higher 
concentration of SNRs and that smoothes the CR distribution in the Disk in comparison 
with that of SNRs. 
Here, we suggest an alternative explanation of this effect.
\subsection{ The radial distribution of HII electron temperature}
It is well known that there is a radial increase of the measured electron temperature
in HII regions with the mean temperature from about 6600$^{\circ}$K at $R = 2kpc$ to 
about 10000$^{\circ}$K at $R = 14kpc$ (\cite{Onire} and Figure 1). This fact arises 
from the 'Galactic Metallicity Gradient' - the fall in metallicity (e.g. Fe/H) with 
increasing $R$ due to the reduced surface density of SNR there (e.g. Figure 1). The 
'metals' act as cooling agents. 

The data on the mean temperature comes from measurements on HII regions and the 
resulting 'warm gas' occupies on average some 30\% of the available volume \cite{Musht}
. However, SN are frequently to be found in regions of OB-associations - and nearby HII
 'objects' - so that most SNR will accelerate background electrons (and protons) having
 'high' temperatures. In what follows we assume that the CR all come from SN exploding 
in regions where the temperature is similar to that in the HII-temperature regions or,
 at least, the similarity is not radially dependent.

\section{The relevance of rising temperature of ISM to the radial distribution of CR 
intensity} 
\subsection{General consideration}
The relevance of rising temperature of the ISM with increasing Galactocentric radius is
 that, as mentioned in \S1, in some CR acceleration models, two of which we examine 
below, the temperature of the ions in the 
ambient ISM is related to the injection efficiency for the ions to take part in the 
acceleration process.The argument is that the particles in the high energy tail 
of the Maxwell-Boltzmann distribution above some 'high' threshold energy 
$\varepsilon_{thr}$ have a number which rises rapidly with temperature when the mean 
energy $< \varepsilon >$ is much less than $\varepsilon_{thr}$. 
Figure 3 illustrates the situation.
\begin{figure}[hptb]
\begin{center}
\includegraphics[width=8cm,height=15cm,angle=-90]{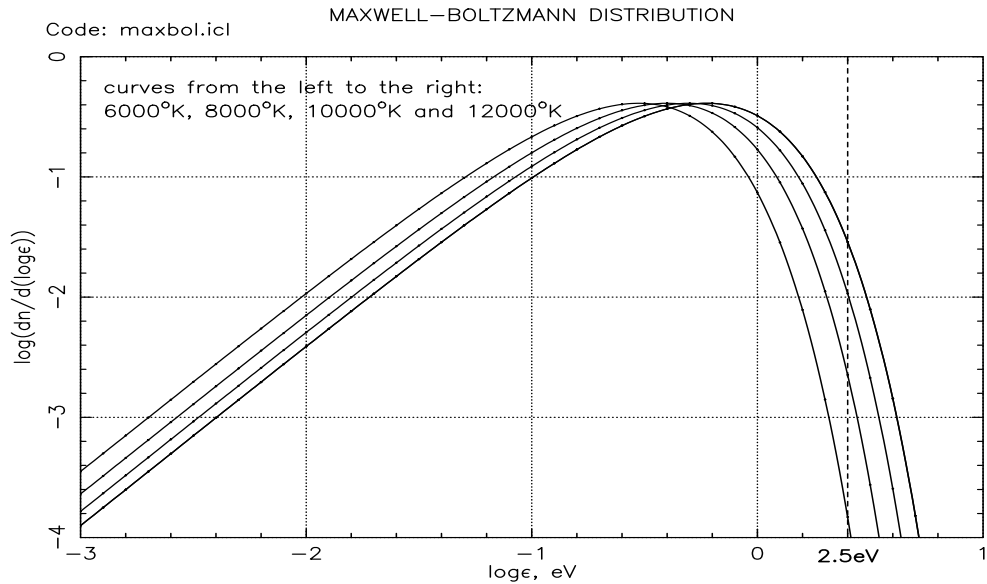}
\end{center}
\caption{\footnotesize Maxwell-Boltzmann distribution for atomic particles}
\label{fig:fig3}
\end{figure}

The inferred CR intensity is proportional to the product of the surface density of 
SNR, $\sigma_{SNR}(R)$, and the efficiency factor $\Delta(R)$, where we assume that 
$\Delta$ is a function of the electron temperature and is the fraction of the 
ionized gas having sufficiently high energy to enter the SNR shock, i.e. the fraction 
with energy greater than $\varepsilon_{thr}$. The problem is to find the necessary 
values of $\varepsilon_{thr}$ and $\Delta$.

In Figure 4 we give $f$ versus $T$ for various values of $\varepsilon_{thr}$. 
Indicated above the abscissa are the limiting values of $R$: 2 and 14kpc to which the 
analysis refers ( the data are too sparse beyond 14kpc and there are complications 
within 2kpc due to black hole effects ).
\begin{figure}[hptb]
\begin{center}
\includegraphics[width=8cm,height=15cm,angle=-90]{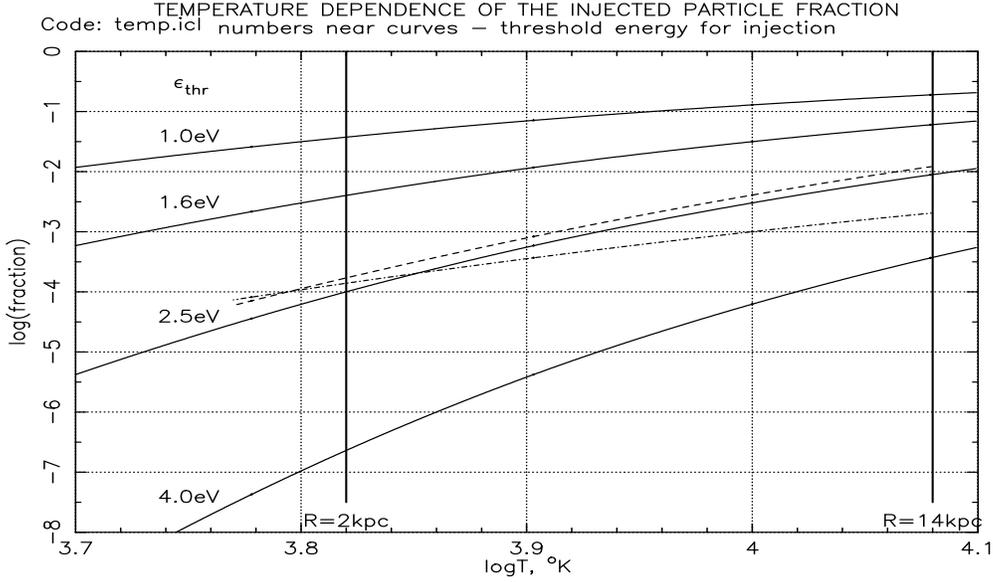}
\end{center}
\caption{\footnotesize Temperature dependence of the injected particle fraction.
$\varepsilon_{thr}$ is the threshold energy for injection. The temperatures at 
$R = 2kpc$ and $14kpc$ from \cite{Onire} are indicated. Fractions of particles injected
from the background plasma and accelerated by the process of the stochastic Fermi 
acceleration or by the SNR shocks are shown by dash-dotted and dashed lines 
respectively (see sections 3.2 and 3.3 below).}
\label{fig:fig4}
\end{figure}

Examining Figures 1 and 2 we see that an increase in efficiency of about a factor 30 is
 needed between $R = 2$ and $R = 14kpc$.  Figure 5 shows the ratio of $\Delta$ vs. 
$\varepsilon_{thr}$ for $R = 14kpc$ and $2kpc$; it shows that a factor of 30 results 
from $\varepsilon_{thr} = 2.5eV$. The corresponding $\Delta$-values are $10^{-4}$ and 
$10^{-2.5}$, and, writing $f \propto T^n$, we have $n \sim 8.4$. It means that the 
fraction of interstellar plasma injected into the acceleration process can rise rapidly
 with the rising temperature.
\begin{figure}
\begin{center}
\includegraphics[width=8cm,height=15cm,angle=-90]{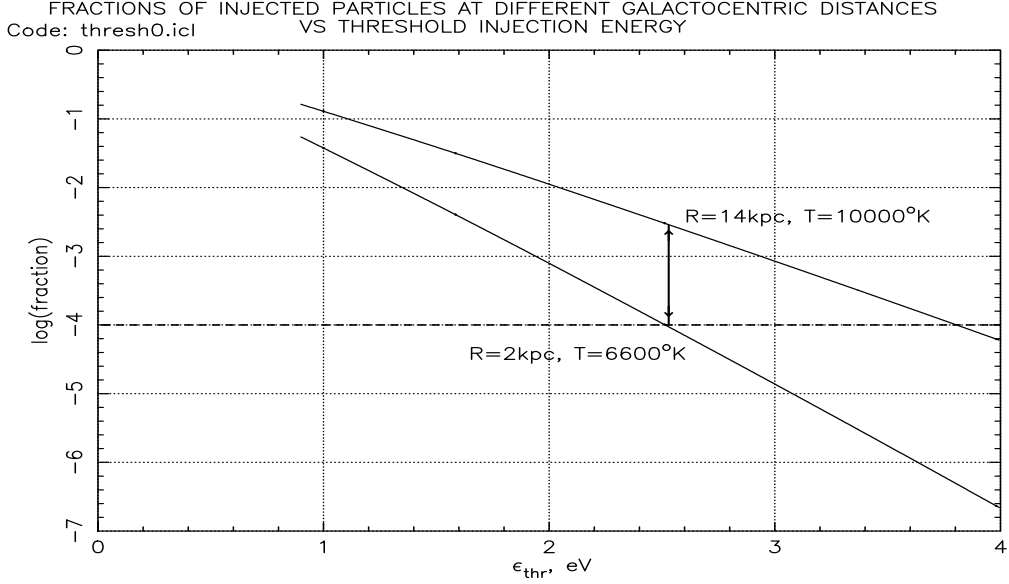}
\end{center}
\caption{\footnotesize Fractions of injected particles at $R = 2kpc$ and $14kpc$ vs. 
the threshold injection energy $\varepsilon_{thr}$.
The value of $\varepsilon_{thr}$ where the ratio of the fractions for $R = 14$ and 
$2kpc$ is equal to 30, necessary to make the CR radial distribution nearly flat, is indicated 
by the vertical double-arrow line.}
\label{fig:fig5}
\end{figure}

We have applied this idea to the known mechanisms of CR production in the Galaxy: the 
stochastic Fermi acceleration and acceleration by SNR shocks. 
 
\subsection{ The process of stochastic CR acceleration from a background plasma}
 The equilibrium Maxwellian momentum spectrum $f_M$ of a background plasma is formed by
 Coulomb collisions. The equation for the distribution function, $f_M$, of background 
plasma with temperature $T$ has a simple form (see \cite{Lifsh})
\begin{equation}
\frac{1}{p^2}\frac{d}{dp}p^2\left[\left(\frac{dp}{dt}\right)_Cf_M+D_C\frac{df_M}{dp}\right]=0
\end{equation}
where $p$ is the particle momentum, $(dp/dt)_C$ is the rate of Coulomb losses, 
$D_C=(dp/dt)_CmkT/p$ is the coefficient of spatial diffusion due to Coulomb collisions,
 $m$ is the mass of the accelerated particles and $k$ is the Boltzmann constant. The 
equilibrium Maxwellian distribution follows, as
\begin{equation}
f_M(p)=\sqrt{\frac{2}{\pi}}n_0\exp\left(-\frac{p^2}{2mkT}\right)=\sqrt{\frac{2}{\pi}}n_0\exp\left(-\frac{E}{kT}\right)
\end{equation}
where $n_0$ is the density of background plasma and $E$ is the particle energy.

If background particles are under the influence of any acceleration, e.g the Fermi 
stochastic acceleration, $(dE/dt)_{ac}=\alpha_0E$, (see \cite{Fermi}), then the energy 
variations of particles in the range $E>kT$ are described by the equation 
\begin{equation}
\frac{dE}{dt}=\left(\frac{dE}{dt}\right)_{ac}-\left(\frac{dE}{dt}\right)_{C}
\label{en_var}
\end{equation}
where $({dE}/{dt})_{C}$ is the rate of ionization losses of a ion with the charge $Z$ and the atomic number $A$ in a gas with the temperature $T$ and the density $n$, which  has the form (see e.g.  \cite{Ginz}) 
\begin{equation}
\left(\frac{dE}{dt}\right)_{C}=\frac{4\pi Z^2 n_0e^4(Am_p)^{1/2}\Lambda}{(kT)^{3/2}m_e}E\left(\frac{E}{kT}\right)^{-3/2}
=Z^2A^{1/2}\nu_0E\left(\frac{E}{kT}\right)^{-3/2}
\end{equation}
where $\Lambda$ is the Coulomb logarithm, $Am_p$ is the mass of the accelerated particles, $m_e$ and $m_p$ are the rest masses of electrons and protons, and   
\begin{equation}
\nu_0=\frac{4\pi n_0e^4m_p^{1/2}\Lambda}{(kT)^{3/2}m_e}
\end{equation}
is the frequency of Coulomb collisions of thermal particles.

For a proton ($Z=1$, $A=1$) Eq. \ref{en_var} has the form
\begin{equation}
\frac{dE}{dt}=\alpha_0E-
\nu_0E\left(\frac{E}{kT}\right)^{-3/2}\label{en_var1}
\end{equation}

From Eq.(\ref{en_var1}) the threshold energy, $\varepsilon_{thr}$ can be derived as
\begin{equation}
\varepsilon_{thr}(n_0,\alpha_0) \simeq kT\left(\frac{\nu_0}{\alpha_0}\right)^{2/3}
\label{eq5}
\end{equation}
which defines the energy range of accelerated particles with $E>\varepsilon_{thr}$.

We notice that the injection energy for a nucleus with mass $Am_p$ , and the charge $Ze$ has the form
\begin{equation}
\varepsilon_{thr}(Z,A) \simeq Z^{4/3}A^{1/3}kT\left(\frac{\nu_0}{\alpha_0}\right)^{2/3}
\label{eq5a}
\end{equation}
As a result, the chemical composition of the accelerated particles differs from that of
 the background plasma (see \cite{Dogi2}). 

The stochastic acceleration forms a power-law spectrum of  nonthermal particles
\begin{equation}
f_{nth}=Kp^{-\gamma}
\end{equation}
and the naive assumption is that the tail of nonthermal particles joins directly to the
 Maxwellian distribution. The constant $K$ in this case is
\begin{equation}
K=\sqrt{\frac{2}{\pi}}n_0\exp\left(-\frac{\varepsilon_{thr}}{kT}\right)
\label{k0}
\end{equation}
and $\gamma$ is the power-law exponent of the spectrum.

However, as \cite{Gurev}  showed, this equation strongly underestimates the number of 
accelerated particles (see also \cite{Dogi3}).  The point is that under the influence 
of acceleration  a flux of particles, running-away into the acceleration region, is 
formed and as a result the particle distribution is nonequilibrium and time variable.  
Because of the running-away flux, the equation for the stochastic Fermi acceleration is
 nonstationary: 
\begin{equation}
\frac{\partial f}{\partial t}-\frac{1}{p^2}\frac{d}{dp}p^2\left[\left(\frac{dp}{dt}\right)_Cf+(D_C+D_F)\frac{df}{dp}\right]=0
\end{equation} 

For the case of weak acceleration, when $\alpha_0<<\nu_0$, the ionization losses have
 a simple form $(dp/dt)_C \propto 1/p^2$ and this equation can be solved analytically. 
This acceleration forms also a power-law spectrum but with the coefficient $K$ being 
different from that in Eq. (\ref{k0}).  For the momentum diffusion e.g. $D_F(p)$, the 
distribution function of accelerated particles is 
\begin{equation}
f(p,t)\sim \sqrt{\frac{2}{\pi}}n_0(t)\exp\left(-\int\limits_{p/\sqrt{mkT}}^\infty\frac{udu}{1+u^3D_F(u)/\nu_0}\right)
\end{equation}
where $n_0(t)$ describes slow temporal variations of the density of background 
particles due to the effect of runaway flux.

In \cite{Gurev} the stochastic acceleration was taken as a power-law, 
$D_F(p)= \alpha_0(p/\sqrt{mkT})^{\varsigma}$. For the Fermi acceleration
  $D_F(p)= \alpha_0(p/\sqrt{mkT})^2$, where $\alpha_0$ is the frequency of 
acceleration. In Fig.4 (dash-dotted line) we present the fraction of 
accelerated particles as a function of the background temperature $T$ for the case of 
stochastic Fermi acceleration with $\varsigma=2$.  
In order to reproduce the fraction of accelerated particles $n_{nth}/n_0\simeq 10^{-4}$
 at $T=6600$ K we take the characteristic acceleration frequency 
$\alpha_0=6\times 10^{-4}$ s$^{-1}$. Here, and below $n_0=10^2$ cm$^{-3}$.

As \cite{Gurev} concluded, the fraction of particles accelerated from a background 
plasma is higher than follows from Eq. (\ref{k0}). The reason is that the acceleration 
generates not only the tail of nonthermal particles but modifies also the component of 
the spectrum which is formed by Coulomb collisions (the Maxwellian component). As a 
result, an extended transfer region is formed between the thermal and nonthermal parts 
of the spectrum.

We notice, however, that this conclusion was derived in the framework of the equation 
when variations of the temperature $T$ are neglected, $T=const$. However, as 
\cite{Wolfe,Petro} showed this effect is not negligible. The point is that the energy 
supplied by sources of stochastic acceleration was quickly absorbed by the thermal 
plasma because of the ionization/Coulomb energy losses of accelerated particles. As a 
result, this acceleration is accompanied mainly by background plasma overheating 
instead of acceleration. 

This conclusion was later revised by \cite{Chern} for the case when the effect of 
acceleration has a cutoff at a low energy of the particles, 
\begin{equation}
D_F(p)= \alpha_0(p/\sqrt{mkT})^\varsigma\theta(p-p_0)
\end{equation}
where $\theta(x)$ is the Heaviside function. This cutoff can be due to the absorption 
of MHD-waves by low energy CRs (see, in this respect, Appendix A in \cite{Cheng}).

In these conditions two competing effects arise: one of them is the plasma overheating which prevents the acceleration; the other is a flux of particles from the thermal pool which cools down the plasma. As  \cite{Chern} showed, for the case when $\varepsilon_{thr}<\varepsilon_0=p_0^2/2m$, acceleration is more effective than the effect of plasma overheating, and the acceleration forms a prominent non-thermal tail. It is interesting to notice that in this case an extended transfer region between the thermal distribution and the tail of accelerated particle (as in \cite{Gurev}) is not formed, and the power-law spectrum joins directly with the Maxwellian distribution at the energy $\varepsilon_0$, that is identical to the estimations shown in Fig.3, where the vertical dashed line 
marks the position of $\varepsilon_0$. 

Estimates of $\varepsilon_0$ is beyond the scope of our analysis. Its value is determined by e.g. spectral characteristics of MGD waves, namely by processes of their excitation and absorption which are not targets of our analysis. For the effective acceleration we need the condition $\varepsilon_0 > \varepsilon_{thr}$ and we ad hoc assume them.

\subsection{The process of CR acceleration by SNR shocks}
The idea of particle acceleration by SNR shocks has been suggested by Krymskii
 \cite{Kryms}, Bell \cite{Bell} and many others. This idea is now confirmed by recent Fermi-LAT 
observations (see e.g. \cite{Thomp}). The Fermi-LAT collaboration 
suggested also that a fraction of CRs is accelerated in OB association and their 
superbubbles \cite{Acke3,Binns}. The medium inside superbubbles is filled by shock 
fronts generated by SN explosions and supersonic winds. In this conditions particles 
are accelerated by interaction with the supersonic turbulence \cite{Bykov} This 
mechanism is similar to the classical Fermi acceleration by magnetic 
fluctuaions/turbulence \cite{Fermi}.

 In the case of CR production by SNRs (that is the goal of this paper), particles 
are accelerated by SNR shocks. In the reference frame of a shock, the process is 
described by the equation (see \cite{Kryms})
\begin{equation}
\frac{\partial}{\partial x}\left(u(x)f-D\frac{\partial f}{\partial x}\right)=-\frac{\nabla {\bf u}}{3}\frac{1}{p^2}\frac{\partial}{\partial p}(p^2f)
 \end{equation}
where the coordinate $x$ is perpendicular to the shock front, $D$ is the coefficient of
 particle spatial diffusion, $u$ is the particle fluid velocity and the velocity jump 
at the shock $(x=0)$ is 
$u_-/u_+=4$ for strong shocks with the Mach number $M>>1$. The characteristic time of 
acceleration in this case is $D/u^2$, and the spectrum of accelerated particles is 
power-law, $f\propto p^{-4}$, or for the energy spectrum 
$N(E)=p^2f(p)dp/dE\propto E^{-2}$. From Eq. (\ref{eq5}) we can estimate the value of 
$\varepsilon_{thr}$ when $\alpha_0=u^2/D$. In order to obtain $n_{nth}/n_0\sim 10^{-4}$
 for $T= 6600$ K we take $u=2\times 10^8$ cm s$^{-1}$ and the spatial diffusion 
coefficient near the shock is $D\simeq 3\times 10^{24}$ cm$^2$s$^{-1}$.

For the case of shock acceleration from a background plasma the kinetic equation is two
 dimensional. As a result, the particle spectrum varies in space, and this makes analysis 
of the acceleration processes quite arduous. This case of acceleration was investigated
 in \cite{Bulan}. The equation has the form
\begin{equation}
\frac{\partial}{\partial x}\left(u(x)f-D\frac{\partial f}{\partial x}\right)-\frac{1}{p^2}\frac{d}{dp}p^2\left[\left(\frac{dp}{dt}\right)_Cf+D_C\frac{df}{dp}\right]=-\frac{\nabla {\bf u}}{3}\frac{1}{p^2}\frac{\partial}{\partial p}(p^2f)
\label{shacc}
 \end{equation}
The solution describes the thermal and nonthermal components of the spectrum and as a 
result it allows us to estimate the number of accelerated particles as a function of 
the parameters of the background plasma. The analysis of Eq. (\ref{shacc}) provided by 
\cite{Bulan} is very bulky. Therefore below we present only the final result for the 
nonthermal component of the  spectrum  at the shock ($x=0$). This spectrum is
\begin{equation}
f_{nth}(x=0,p)=C(n_0,T)p^{-4}
\end{equation}
From this equation we can derive the ratio $n_{nth}/n_0$:
\begin{equation}
\frac{n_{nth}}{n_0}\simeq \frac{p_T}{p_{thr}}\left(\frac{m_e}{m_p}\right)\delta^{1/3}
\exp\left\{-\delta^{1/2}\left(1+\frac{1}{2}\ln\left[\left(\frac{m_i}{m_e}\right)^2\right]\frac{1}{3\delta}\right)\right\}
\label{ratio}
\end{equation}
where $m_e$ and $m_p$ are the electron and proton mass respectively, $m_i$ is the mass 
of the accelerated particles,  
\begin{eqnarray}
&&p_T=\sqrt{2kTm_i}\nonumber\\
&&p_{thr}=p_T\sqrt{\frac{m_p}{m_e}}\delta^{1/3}\nonumber\\
&&\delta=\frac{D\bar{\nu}}{u^2}\nonumber\\
&&\bar{\nu}=\frac{2\pi n_0e^4m_p^2}{m_e\bar{p}^3}\Lambda\nonumber\\
&&\bar{p}=m_p\sqrt{\frac{2kT}{m_e}}\nonumber
\end{eqnarray}
The ratio $n_{nth}/n_0$ calculated from Eq. (\ref{ratio}) is shown in Fig.4
( dashed line ). 
In the standard models of CR propagation the term describing CR injection by SNRs  is 
supposed to be proportional to the density of SNRs, $N_{SNR}(R)$, at the Galactocentric
 radius $R$. However, as one can see from Fig.4 for the case of shock 
acceleration from a background plasma it depends also on the temperature of the 
background plasma.  
\section{Discussion and conclusions}
We see that both in our general consideration and in more sophisticated models the 
fraction of background particles injected into the acceleration process rises with the 
ambient ISM temperature. In first order, the values of $\varepsilon_{thr} = 2.5eV$ and 
$10^{-4}/10^{-2.5}$ for
the fractions appear 'reasonable' but there are a number of uncertainties. The most 
important concerns the extent to which the gas densities in the regions in which CR are
 accelerated are proportional to the total gas densities in those regions and the 
temperatures likewise. If they are then the value of $\varepsilon_{thr}$ will be 
appropriate and the $\Delta$-values can be regarded as injection efficiencies and can 
be compared with the values commonly adopted; for example, the values adopted as 
examples in \cite{Berez} range from $10^{-2}$ to $10^{-4}$, i.e. in our range. Writing 
$\Delta \propto T^n$ we have $n = 8.4\pm 3.2$. 

In conclusion, we have presented a model in which the gradient of electron temperatures
 in the Galaxy can make a contribution, at least, to solving the  cosmic ray gradient 
problem. A proper evaluation of the implications of the background particle temperature effects
for the CR gradient problem will require a more detailed analysis taking into account the 
acceleration and propagation processes in a self-consistent manner.

{\bf Acknowledgements}

ADE and AWW are grateful to the Kohn Foundation for financial support. VAD acknowledges
partial support from the RFFI grants 15-52-52004, 15-02-92358 and from the 
International Space Science Institute to the International Team 'New Approach to the 
Active Processes in Central Regions of Galaxies'.

\end{document}